\documentclass[%
 reprint,
superscriptaddress,
 amsmath,amssymb,
 aps,
prl,
]{revtex4-2}

\usepackage[greek,english]{babel} 
\usepackage[OT2,OT1]{fontenc}
\usepackage{svg}
\svgsetup{
    inkscapepath=i/svg-inkscape/
}
\svgpath{{svg/}}
\usepackage{graphicx}
\usepackage{dcolumn}
\usepackage{bm}
\usepackage{oubraces}
\usepackage{xspace}
\newcommand{\Poincare}{Poincar\'e\xspace}
\newcommand{\upcite}[1]{\textsuperscript{\textsuperscript{\cite{#1}}}}

\begin{document}

\preprint{APS/123-QED}

\title{Spherically Symmetric Noncommutative Spacetime via Exotic Atomic Transitions}

\author{Junlin Wu}
\author{Horan Tsui}
 \thanks{Junlin Wu and Horan Tsui contributed equally to this work}
\author{Bowen Tong}
\author{Shin-Ted Lin}
\altaffiliation{Corresponding author}
\email{\\stlin@scu.edu.cn}
\author{\\Shu-Kui Liu}
\affiliation{%
 College of Physics, Sichuan University, Chengdu 610064 
}%
\author{Muhammed Deniz}%
 \altaffiliation{Corresponding author}
 \email{\\muhammed.deniz@deu.edu.tr}
\affiliation{
 Department of Physics, Dokuz Eyl\"ul University, Buca, \.Izmir TR35390
}%
\author{Henry T. Wong}
\affiliation{
Institute of Physics, Academia Sinica, Taipei 11529
}
\author{Qian Yue}
\affiliation{
Key Laboratory of Particle and Radiation Imaging, Tsinghua University, Ministry of Education, Beijing 100084
}

\begin{abstract}
In discussing noncommutative spacetime, the generally studied $\theta$-\Poincare model is inconsistent with bound states. In this Letter, we develop the formalism and study the phenomenology of another model $\mathcal{B}_{\chi \hat{n}}$ by the twisted permutation algebra and extend the Pauli Exclusion Principle (PEP) into noncommutative spacetime. The model also implies time quantization and can avoid UV/IR mixing. Applying it to atomic systems, we show that the model with newly induced phase factors can cause exotic transitions consisting of three electrons in the 1S orbit of atoms. The transition rate is derived, and the upper bound of noncommutative parameter $\chi$ is thus set by utilizing data from the low-energy and low-background experiments, where strongest constraint $\chi\leq4.05\times10^{-30}$ eV$^{-1}$ at 90\% C.L. is given by XENONnT, with the time quanta $\Delta t\sim 2.67\times 10^{-45} s$, equivalent to twenty times smaller than the Planck time.
\end{abstract}

\maketitle


The short distance singularity would be eliminated by quantum fluctuation if spacetime is noncommutative:  $[x^\mu,x^\nu]\neq 0$. Therefore, it is widely accepted that noncommutative spacetime will appear, at least as an effective description, in many quantum gravity theories, including the string theory.

The first attempt goes back to W. Heisenberg \cite{H1}. And then R. Peierls \cite{peierls33} has shown that noncommutative spacetime can produce the Landau level. H. Snyder \cite{Snyder:1946qz} first published a well-known work which was followed by C. N. Yang $\textit{et al}$ \cite{Yang47}. An actual sensation comes after the discovery made by A. Connes $\textit{et al}$  \cite{Connes:1997cr} that the BFSS matrix model \cite{Banks:1996vh} and the IKKT matrix model \cite{Ishibashi:1996xs} can be bridged by noncommutative geometry. Based on this work, N. Seiberg and E. Witten \cite{Seiberg:1999vs} have proved that the open-string dynamics can be described by a noncommutative gauge theory in some limits. 

The mathematical frame for noncommutative spacetime $\mathbb{R}^{d+1}$ is the twisted Hopf algebra on it, replacing the regular product $``\cdot"$ with the Moyal $``\star"$ product \cite{moyalplaneBalachandran:2005eb,moyalplane3Akofor:2008ae}: 
\begin{align}
    f(x)\star g(x):&= F f(x+\eta) g(x+\eta)|_{\eta \rightarrow 0}
\end{align}
where $f(x), g(x)$ are functions (operators) on $\mathbb{R}^{d+1}$, and $F$ is Drinfel'd twist \cite{Drinfeld:1986in}, deforming one Hopf algebra into a corresponding noncommutative one.

The simplest Lorentz covariant model is $\theta$-\Poincare model \cite{theta1Chaichian:2004za,theta2Addazi:2018jmt,strongestpepv}, setting $F=e^{\frac{i}{2}\theta^{\mu\nu}\partial_\mu\partial_\nu}$ ( $\hbar=c=1$ ), where $\theta_{\mu\nu}$ is an antisymmetric tensor, and $[x^\mu, x^\nu]=i \theta^{\mu\nu}$. However, except for simplicity, it has however some conceptual confusions, like UV/IR mixing \cite {Minwalla_2000, Craig_2020, UVIR3}, and technical obstacles to calculate with, especially for bound states whose energy eigenstates will generally be broken by linear momentum operators $-i\partial_{\mu}$. 

Therefore, we adopt another model, $\mathcal{B}_{\chi \hat{n}}$, proposed by A. P. Balachandran and P. Padmanabhan \cite{pepv2Balachandran:2010xk,paperBalachandran:2010wq} with $F=e^{-\frac{i}{2}  \chi (P_0 \otimes J_{\hat{n}}-J_{\hat{n}}\otimes P_0)}=e^{-\frac{i}{2}\chi (P_0 \wedge J_{\hat{n}}) }$, and   
\begin{align}
[x^0,x^i]&:= x^0 \star x^i- x^i \star x^0= i\chi\epsilon_{kij}n^k x^j \nonumber\\
[x^i,x^j]&=0 \hspace{1cm}  i,j,k=1,2,3
\end{align}
where $[\chi]=M^{-1}$ is the model parameter measuring the non-commutativity, which should be small for the consistency with experiments, $P_0=-i\partial_0$, and $J_{\hat{n}}=-i \epsilon_{ijk}\hat{n}^i \hat{x}^j \partial_k $ is the angular momentum operator along $\hat{n}$. The commutator can also be obtained by equating $x^0$ with $\chi J_{\hat{n}}$, which amounts to time quantization in the unit of $\chi$. Note that $\epsilon_{ijk}$ in Eq. 2 implies a cyclic uncertainty relation, from which we also have the minimum time unit 
\begin{align}
    \Delta t= \chi
\end{align}

The model essentially quantizes only one coordinate and imposes nothing on space coordinates, and has no UV/IR mixing. It also protects the energy eigenstates of rotationally symmetric systems, like atomic systems. For such system of well-defined angular momentum quantum numbers, $\hat{n}$ plays the same role as $\hat{z}$, the conventional projection axis, and $J_{\hat{n}}\to J_{\hat{z}}$. One should simultaneously vary it when the systems are dynamic or inhomogeneous. 

The break of regular spacetime structure will twist the spin-statistics \cite{PEPVst2,PEPVst3,PEPVst4} and further cause one of the most salient phenomena, the Pauli Exclusion Principle Violation (PEPV)  \cite{moyalplaneBalachandran:2005eb,pepv1Greenberg:1987ih,pepv5Ramberg:1988iu,pepv6VIP:2006gvw,pepv7Borexino:2004hfc}. Specifically, in $\mathcal{B}_{\chi \hat{n}}$ model, as we will show, the newly induced phase factors from $F$ blur the definition of and difference between the fermions and bosons, and the new particles are denoted as twisted fermions and bosons as the convention. 

For the experimental detection, we focus on the exotic transition whose final state consists of three electrons in 1S of atoms. The amplitude for this process is the function of $\chi$. Therefore one can set the upper bound of $\chi$ by comparing it with experimental observation. However, what is subtle is that electrons in the 1S orbit are twisted bosons rather than twisted fermions since, as we will show, the Pauli Exclusion Principle (PEP) can be partly extended into noncommutative spacetime by taking the phase factors into account, ruling out identical twisted fermions.

$\textit{Permutation Algebra.---}$In the field theory of regular spacetime, the N-particle states of Fock space are produced via the series of creation and annihilation operators 
\begin{align}
    &|p_1,p_2,...p_N\rangle=a_{p1}^{\dagger}\cdot a_{p2}^{\dagger}\cdot... a_{pN}^{\dagger}|0\rangle\nonumber\\
     =&[m(m\otimes I)...(m\otimes\underbrace
    {I\otimes ...I}_{N-2})]a_{p1}^{\dagger}\otimes a_{p2}^{\dagger}\otimes... a_{pN}^{\dagger}|0\rangle
\end{align}
where $m$ is the product map: $m(a\otimes b)=a\cdot b$, and ``$\cdot$" is the regular product between operators. Swapping N operators requires permutation algebra $\mathbb{C}G$ \cite{Hazewinkel10, Majid1995}. Consider one of its elements, cyclic permutation $\sigma_{c}$ for illustration
\begin{align}
&a_{pN}^{\dagger}\otimes a_{p1}^{\dagger}\otimes ... a_{pN-1}^{\dagger}\nonumber\\
=&\eta^N [(\underbrace{I\otimes ...I}_{N-2}\otimes\Delta)...(I\otimes\Delta)\Delta\sigma_{c}]a_{p1}^{\dagger}\otimes a_{p2}^{\dagger}\otimes... a_{pN}^{\dagger}
\end{align}
where $\eta=1$ for bosons and -1 for fermions, and $\Delta$ is the co-product map: $ \Delta(\sigma_{{\{n\}}})=\sigma_{{\{n\}}}\otimes\sigma_{{\{n\}}}$, where ${\{n\}}$ is the permutation sequence. 

$\textit{Twisted Atomic Systems.---}$Since then, we focus on atomic systems and assume all states are the eigenstates of $P_0$ and $J_{\hat{n}}$. In noncommutative spacetime, $m$ and $\Delta$ are both changed by the twist $F=e^{-\frac{i}{2}\chi (P_0 \wedge J_{\hat{n}}) }$
\begin{align}
m&\to m_F(f\otimes g):=m[F (f\otimes g)]=f\star g\\
    \Delta&\to\Delta_F:=F\Delta F^{-1}
\end{align}
Substituting them into Eqs. 4 and 5 yields the twisted Fock space. To study the energy of N-particle states after twisting, we should extrapolate the operator $P_0$ as Eq. 5 by $\Delta$, and as an element of $\mathbb{C}P$, a Lie algebra, $\Delta$ acts as $\Delta P_{\mu}=P_{\mu}\otimes I+I\otimes P_{\mu}$ \cite{Weinzierl03}. Since 
\begin{align}
[F=e^{-\frac{i}{2}\chi(P_0\wedge J_{\hat{n}})},P_0]\sim -\frac{i}{2}\chi P_0\wedge[J_{\hat{n}},P_0]
\end{align}
the energy remains the same rigorously for systems rotationally invariant along $\hat{n}$ given $[P_0, J_{\hat{n}}]=0$, and approximately for other systems given $\chi\ll 1$. 

Then consider the effect of $F$ on such systems
\begin{align}
& a^\dagger_{p1}\star  a^\dagger_{p2}|0\rangle\nonumber\\
   =&m[\sum_n \frac{1}{n!}(-\frac{i}{2}  \chi)^n (P_0 \otimes J_{\hat{n}}-J_{\hat{n}}\otimes P_0)^n|p_1\rangle\otimes|p_2\rangle]\nonumber\\
    =&e^{-\frac{i}{2}\chi (E_1 J_2-J_2 E_1)}|p_1,p_2\rangle:=F(1,2)|p_1,p_2\rangle
\end{align}
where $E_i, J_i$ denote the eigenvalues of $|p_1\rangle$ and $|p_2\rangle$. It is these extra state-dependent phase factors in noncommutative spacetime that twists states. Note that by the form of $\Delta_F$, swapping now not only interchanges particles but also alters the phase factors correspondingly. For any sequence of N-particle states (omit the subscripts of $\sigma_{\{n\}}$ for simplicity)  
\begin{align}
  &|\sigma(p_1),\sigma(p_2)...\sigma(p_N)\rangle\nonumber\\
  \to& \amalg_{i<j;1}^N F[\sigma(i),\sigma(j)]|\sigma(p_1),\sigma(p_2)...\sigma(p_N)\rangle
  \end{align}It is now straightforward to rule out the identical twisted fermions. Suppose fermions $j$ and $k$ are identical, that is $E_j=E_k, J_j=J_k$. Swapping them will produce a minus sign ``-" but not affect the phase factor $\amalg_{l<m} F(l,m)$ that depends only on $E$ and $J$, hence
\begin{align}
    |\phi\rangle_{fermion}=- |\phi\rangle_{fermion}&\equiv 0
\end{align}

The state is forbidden. It is the generalization of the Pauli Exclusion Principle (PEP) in noncommutative spacetime in spherically symmetrical systems. Note that it is true for not only the $\mathcal{B}_{\chi\hat{n}}$ but also any models in the scenario of Drinfel'd twist. 
\begin{figure}[b]
\includegraphics[width=1\linewidth]{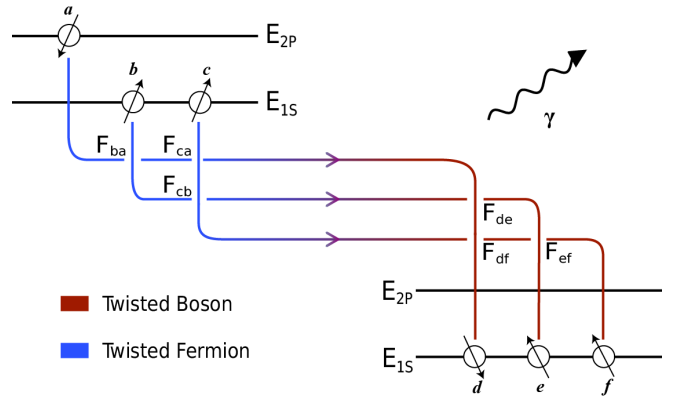}
    \caption{Transition between a twisted fermionic state $F_{ba}F_{ca}F_{cb}|c,b,a\rangle_{\hat{n}}$ to a twisted bosonic state $F_{de}F_{df}F_{ef}|d,e,f\rangle_{\hat{m}}$ with spin arrows and an emitted photon. The overlap of strings denotes the sequence, for example, the string of electron $d$ is above that of $e$ and $f$, hence the state $|d,e,f\rangle_{\hat{m}}$ and the phase factors $F_{de}$, $F_{df}$ at the corresponding junctions. Swapping the upper and lower strings will change the factors and produce a minus (plus) sign for the blue (red). When $\chi\to 0$, all $F\to 1$, the twisted becomes the standard, and such transition is forbidden.}
\end{figure}

For the further calculation of phase-factor-involved transition amplitudes, we expand the state by a common trick of adding up all possible permutations of the Fock space state  
\begin{align}
&\amalg_{i<j,1}^N F(i,j)|p_1,p_2...p_N\rangle\nonumber\\
= & \frac{1}{N!}\sum_{\{n\}}\amalg_{i<j,1}^{N}F[\sigma(i),\sigma(j)]|\sigma(p_1), \sigma(p_2),...\sigma(p_N)\rangle
\end{align}
which will recover the familiar expansion when $\chi\to 0$, and the phase factor in Eq. 12 is different from the previous work \cite{paperBalachandran:2010wq}, which additionally has $\amalg_{i<j,1}^{N} F^{-1}(i,j)$. 

Now consider three electrons in a given atomic system, set two of them in 1S orbit and another in an allowed higher orbit $X$, and treat everything else as the background by the mean-field approximation. By Eq. 12, the twisted fermionic initial state of three electrons is 
\begin{widetext}
\begin{align}
   3!|\psi\rangle=&e^{-\frac{i}{2}\chi   E_{1s}(2J_X-1)}|1+,1-,X\rangle-e^{-\frac{i}{2}\chi E_{1s}(-2J_X+1)}|X,1-,1+\rangle+e^{-\frac{i}{2}\chi E_{1s}(-1-2J_X)}|X,1+,1-\rangle\nonumber\\
   -& e^{-\frac{i}{2}\chi(-E_{1s}-E_X)}|1+,X,1-\rangle+e^{-\frac{i}{2}\chi( E_{1s}+E_X)}|1-,X,1
    +\rangle-e^{-\frac{i}{2}\chi E_{1s}(1+2J_X)}|1-,1+,X\rangle
\end{align}
\end{widetext}
where in ``kets" the numbers  denote the orbit occupied and ``$\pm$" represent spin up and down. We have set $J_{1s}=\frac{1}{2},-\frac{1}{2}$ for $1+$ and $1-$ respectively. 

We can further set $X$ = 2P since 2P - 1S is the dominant process among all transitions of the final state in 1S. Average 2P$_{1/2}$ and 2P$_{3/2}$ for experimental convenience 
\begin{align}
    |X\rangle:= & \frac{1}{6}\sum_{l=\pm 1,0}\sum_{s=\pm\frac{1}{2}}|l, s\rangle\nonumber\\
    E_X :=&\frac{1}{2}(E_{2P_{1/2}}+E_{2P_{3/2}})
\end{align}
The final state, however, cannot be the twisted fermionic state due to Eq. 11. It is also straightforward to check by setting $X$ = 1S ($J_X=\pm\frac{1}{2}$) in Eq. 13. Nonetheless, there is a non-vanishing twisted bosonic final state, turning all ``-" into ``+", for three electrons, which has four possible configurations being the eigenstates of $J_{\hat{n}}=\pm\frac{3}{2}, \pm\frac{1}{2}$ respectively. We sum four configurations as the final state
\begin{align}
       |\phi\rangle:=\sum_{s=\pm\frac{3}{2},\pm\frac{1}{2}}|\phi,s\rangle
\end{align}
$\textit{Transition Amplitude.---}$Denote the amplitude for such an exotic transition as $A_{\chi}$. 

Note that in regular spacetime the transition is forbidden due to the spin-statistics theorem vanishing the final state rather than the disability of the interaction. Therefore, in noncommutative spacetime we still adopt the standard interaction $V$ causing the 2P - 1S transition for a single electron. Omit electron correlations since they are small and irrelevant to the transition, the interaction for three-electron twisted Fock states is $ (I\otimes \Delta_F)\Delta_F (V)$, which involves many phase factors. Since $V$ is generally not spherically symmetric, the correction $\delta V\sim \mathcal{O}(\chi)$ due to Eq. 8. However, as we will show, $A_{\chi}\sim \chi$ without considering $\delta V$, thus the amendment $\delta A_{\chi}\sim \mathcal{O}(\chi^2)$ is negligible, hence the interaction $\sim I\otimes I\otimes V+I\otimes V\otimes I+V\otimes I\otimes I$.

\begin{table*}[t]
\caption{\label{tab:table1}
Summary of the experimental limits on $\chi$. The exposure volume is given considering efficiency cuts, which are illustrated respectively by reference. The classification of the statistical analyses is described in the text.}
\begin{ruledtabular}
\begin{tabular}{ccccccc}
 Atom&Experiment&Observed Energy\footnote{The observed energy $E_{1s}$ is derived from Ref. \cite{t2}.}&Exposure Volume&$\rm Chi\ Square/\rm ndf$&Exotic Count Rate&$\chi\ \text{Limit}$ \\
 &&[keV]&[kg day]&&(90\% C.L.)[events/(kg d)]&(90\% C.L.)[eV$^{-1}$] \\\hline
 Ne&NEWS-G&0.870&7.71 \upcite{ng2}&14.63/24&$5.41$  $^{\rm (\romannumeral3)}$&$1.17\times10^{-24}$ \\
 Si&DAMIC&1.839&9.90 \upcite{dami2}&26.22/24&$1.35$  $^{\rm (\romannumeral2)}$&$1.39\times10^{-25}$\\
 Ar&DarkSide-50&3.206 (49.59 e$^-$)&12665.57 \upcite{dk2}&77.03/49&$2.02\times10^{-3}$ $^{\rm (\romannumeral3)}$&$2.06\times10^{-27}$\\
 Ca&CRESST-\uppercase\expandafter{\romannumeral2}&4.039&2.75 \upcite{cr4}&121.8/111&$2.99\times10^{-1}$ $^{\rm (\romannumeral2)}$&$1.58\times10^{-26}$\\
 Ge&MJD&11.103&12480.19 \upcite{mjd2}&7.38/31&$7.83\times10^{-4}$ $^{\rm (\romannumeral1)}$&$1.69\times10^{-28}$\\
 I&DAMA/LIBRA\footnote{We choose an example of the energy spectrum of the single-hit scintillation events collected by one DAMA/LIBRA-phase2 detector in one annual cycle, as the typical spectrum in all annual cycles of DAMA/LIBRA-phase2 (Nov. 2, 2011 — Sept. 25, 2017).}&33.169&411137 \upcite{dama2}&Unidentified \footnote{The limit is derived by the exposure volume and detector resolution, along with the indistinguishable quantities between data and background model.}&$5.86\times10^{-3}$ $^{\rm (\romannumeral3)}$&$6.14\times10^{-29}$\\
 Xe&XENONnT&34.561&338720 \upcite{xe4}&21.57/29&$2.88\times10^{-5}$ $^{\rm (\romannumeral2)}$&$4.05\times10^{-30}$\\
 W&CRESST-\uppercase\expandafter{\romannumeral2}&69.525&12.93 \upcite{cr4}&71/58&$1.79\times10^{-1}$ $^{\rm (\romannumeral2)}$&$1.00\times10^{-28}$\\
\end{tabular}
\end{ruledtabular}

\end{table*}

With above all, the total amplitude is 
\begin{align}
&(3!)^2 A_{\chi}(\hat{m},\hat{n})={}_{\hat{m}}\langle \phi |(I\otimes\Delta_F)\Delta_F(V)|\psi\rangle_{\hat{n}}\nonumber\\
    \sim& \frac{1}{2}\sum_{s_{\psi}}\sum_{s_{\phi}}{}_{\hat{m}}\langle s_{\phi}|s_{\psi}\rangle_{\hat{n}}F^{-1}_{\phi}F_{\psi}\frac{1}{3}\sum_{l=\pm1,0}\langle 1|V|l\rangle
\end{align}
where $|s_{\psi}\rangle_{\hat{n}}$ and $|s_{\phi}\rangle_{\hat{m}}$ are the three-particle spin configurations for the initial and the final states respectively. A typical term of $|s_{\psi}\rangle_{\hat{n}}$ is $|+,-,s_X\rangle_{\hat{n}}$. The selection rule, $\langle 1|V|1\rangle=0$, has been used. By Eqs. 14 and 15, denote $\frac{1}{3}\sum_{l}\langle 1|V|l\rangle$ as $A_0$, the polarization-free amplitude for the regular single electron transition which can be detected by experiments. 

\begin{figure}[b]
\includegraphics[width=1\linewidth]{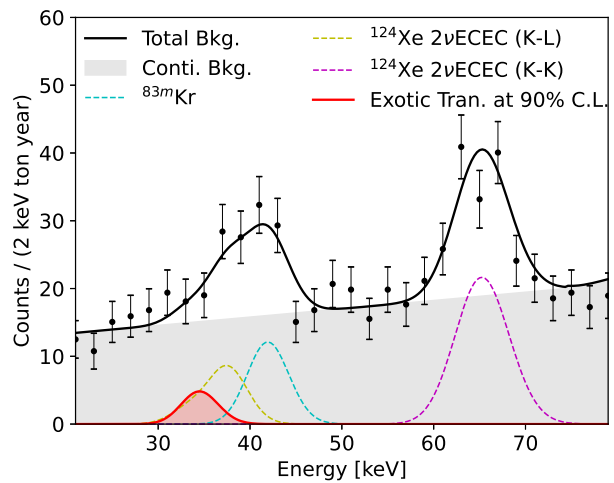}
\caption{\label{fig:epsart2}Data from XENON-nT and the best-fit with the background model \cite{xe4} and its peak signature. A 90\% C.L. upper limit on the peak signature labeled in red is superimposed.}
\end{figure}

Note that generally the symmetric direction of the final state can differ from the initial. After integrating the final direction as Ref. \cite{paperBalachandran:2010wq}, the amplitude is
\begin{align}
    A_\chi=\int A_{\chi}(\hat{m},\hat{n})\frac{d\Omega_{\hat{m}}}{4\pi}=i g_\chi A_0
\end{align}
where 
\begin{align}
    g_\chi=\frac{11}{9}\text{sin}(\chi E_{1s})+...\sim\frac{17}{6}\chi (E_{1S}-E_X)
\end{align}
is the square root of the branching ratio of the exotic transition to the regular transition, and $\frac{17}{6}$ is solely dependent on the angular momentum configuration.

$\textit{Experimental Constraints.---}$Experimentally, an exotic transition can be observed by measuring x-ray emissions with particular energy generated by this transition in a given atomic structure. It can be achieved by scanning for exotic signal peaks in selected regions of electron recoil energy spectra, derived from some rare event search experiments which focus on low-energy regions. In this letter, we give 90\% C.L. constraints on $\chi$ , through the data from M\scriptsize{AJORANA} \normalsize{D}\scriptsize{EMONSTATOR}\normalsize{} (MJD) \cite{mjd2}, XENONnT \cite{xe4}, DAMA/LIBRA \cite{dama1}, DAMIC \cite{dami2}, CRESST-II \cite{cr3,cr4}, NEWS-G \cite{ng2}, DarkSide-50 \cite{dk2}.

The energy released in the exotic transition 2P - 1S should be the difference of binding energy $E_{1S}-E_X$. However, other consequent regular processes would soon come after the exotic transition till the atomic system reaches equilibrium. Therefore, the observed energy should be the sum of all eigenenergy differences equal to the binding energy of the final state (K-shell) $E_{1S}$. In a stable system, it's hard for K-shell electrons to be excited and to generate any vacancies. Thus signals with 1S binding energy must be produced by exotic transitions.

In data analysis, a minimum chi square analysis is applied to the background model, weighting the counts by analytic cut efficiency. We use a free (two free) parameter(s) characterizing the background level and the possible $\chi$ value. For most experiments, signal peaks are pure Gaussian. In the analysis of XENONnT, a skew-Gaussian smearing function is considered \cite{xe5}. For DarkSide-50, we use the calibration signal shape of $^{37}$Ar K-shell electron capture\cite{dk4} as the shape for Ar exotic signal, thanks to the same e$^-$. The LAr ionization response to electronic recoils is given by Ref. \cite{dk4}.

\begin{figure}[h]
\includegraphics[width=1\linewidth]{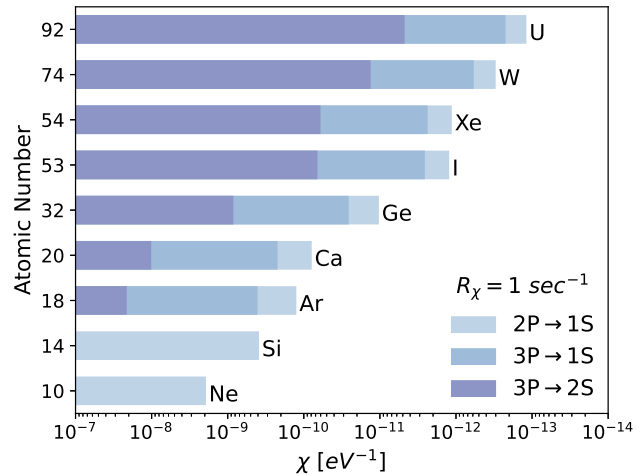}
\caption{\label{fig:epsart1} Comparison of the $\chi$ sensitivities with $R_{\chi}$ set to unity and various atomic species at different exotic atomic transitions. }
\end{figure}

The energy resolution is given by three different methods. In Method.(\romannumeral1), using fitted peak shapes, the energy resolution \cite{mjd2} is ${{\sigma}_{D}(E) = \sqrt{{\Gamma^2_n}+{\Gamma^2_F}E+{\Gamma^2_q}E^2}}$, where free parameters $\Gamma_n$, $\Gamma_F$, and $\Gamma_q$ account respectively for electronic noise, the Fano factor \cite{t4}, and incomplete charge collection. In Method.(\romannumeral2), the energy resolution is given by ${\sigma}_{D}(E)=a+b\sqrt{E}$, where the free parameters $a$, $b$ are fitted using known peaks. In Method.(\romannumeral3), the resolution is derived from corresponding references directly. The required parameters are shown in Table~\ref{tab:table1}. The best fit for XENONnT is shown in Fig.~\ref{fig:epsart2} as an example. 

\begin{figure}[b]
\includegraphics[width=1\linewidth]{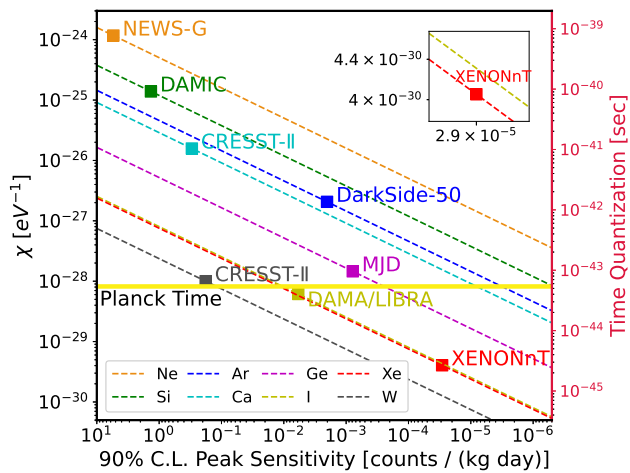}
\caption{\label{fig:epsart} 
90\% C.L. upper limits on $\chi$, reflected in time quantization unit, are placed from experiments. Dash lines depict the relation between the search of $\chi$ with different target atoms and the peak signature sensitivity associated with the exposure volume, peak resolution, and background levels.
}
\end{figure}

The 90\% C.L. upper limits on the exotic transition rates are given by the formula
\begin{align}
    R_{\chi}\leq\frac{N_{U}}{N_{N}N_{n}T}
\end{align}
where $N_{U}$ is the 90\% C.L. counts of exotic transitions derived via the Feldman-Cousins procedure \cite{t3}, $N_{N}$ is the total number of atoms under consideration, $N_{n}$ is the number of electrons in the initial state where the exotic transitions are possible, and $T$ is the time of measurements. By Eqs. 17 and 18, we have
\begin{align}
    \chi\leq\frac{17}{6(E_{1S}-E_X)}\sqrt{\frac{R_{\chi}}{R_{0}}}
\end{align}
where $R_{0}$ is the regular transition rate weighted average rate of two processes 2P$_{1/2}$ - 1S$_{1/2}$ and 2P$_{3/2}$ - 1S$_{1/2}$, derived from the relativistic Hartree-Slater calculation by J. H. Scofield \cite{t5}, and $R_{\chi}/R_{0}=|A_{\chi}/A_0|^2$ is regarded as the branching ratio. The results are also shown in Table~\ref{tab:table1}. Fig.~\ref{fig:epsart1} shows that the atoms with a larger atomic number provide higher sensitivity in the search for exotic transitions because of larger energy differences and regular transition rates. Therefore, uranium has the maximum probability to generate exotic transitions among naturally occurring elements. 

$\textit{Conclusions.---}$We have shown in the bound-state consistent model $\mathcal{B}_{\chi\hat{n}}$, the noncommutative effect comes down to extra phase factors, enabling exotic twisted fermion-boson transition generally with the amplitude $\sim\chi$. One can further prove that the amplitude of twisted fermion(boson)-fermion(boson) transitions are almost irrelevant to $\chi$. A complete understanding of twisted bosonic states of electrons may require quantum gravity. Nonetheless, the non-zero overlap Eq. 18 attests to its existence given the nature of quantum that everything possible finally happens.

Fig.~\ref{fig:epsart} shows that the data from XENONnT gives the strongest constraint $\chi\leq 4.05\times 10^{-30}$ eV$^{-1}$ or $\frac{1}{\chi}\geq 20\times \text{Planck Mass}$. Although atoms with a larger atomic number are more likely to generate exotic transitions, we think that the LXe projects would be the most ideal type of experiments for exotic transitions searching, for their large exposure and good sensitivity. 

The time quanta $> 0.05 \times$ Planck Time is the unique prediction of $\mathcal{B}_{\chi \hat{n}}$ model. It essentially originates from involving $J_{\hat{n}}$ in the twist, and thus any noncommutativity models consistent with rotational systems may share this trait. The results which we obtained shorter than Planck Time favor the idea that no currently available physical theories can describe such a short time where current physical approaches and understandings are broken down. The limits shown in Fig.~\ref{fig:epsart} indicate the time shorter than Planck Scale is possible and exists, therefore we believe our work implies strongly that new theories of quantum gravity are necessary and open a new window in those fields via noncommutative spacetime \cite{report,timeq1}.  

Besides, it is possible to connect $\mathcal{B}_{\chi \hat{n}}$ with $\theta$-\Poincare by re-expressing the twist:  $F=e^{-\frac{i}{2}P_0\wedge J_{\hat{n}}}\to e^{-\frac{i}{2}\theta_{abc}P^a P^b x^c}$, where $\theta_{abc}$ in this case is $\theta_{0ij}\sim\lambda_{0b}\epsilon_{bij}$, and $\lambda_{0b}$ is determined by $\hat{n}_b$.

Bound states are important in constructing our world, a systematic study of noncommutativity effect on it is thus indispensable. 

This work is supported by the National Natural Science Foundation of China (Grants No. 11975159, No. 119775162) and the SPARK project from the research and innovation program of Sichuan University [2018SCUH0051]. The authors are grateful to Prof. Chun-Lin Bai, Prof. Ming Lu, and Prof. Ekrem Ayd$\i$ner for the useful discussion and comments.

\nocite{*}
\bibliography{ms}

\providecommand{\noopsort}[1]{}\providecommand{\singleletter}[1]{#1}%
\begin{thebibliography}{46}%
\makeatletter
\providecommand \@ifxundefined [1]{%
 \@ifx{#1\undefined}
}%
\providecommand \@ifnum [1]{%
 \ifnum #1\expandafter \@firstoftwo
 \else \expandafter \@secondoftwo
 \fi
}%
\providecommand \@ifx [1]{%
 \ifx #1\expandafter \@firstoftwo
 \else \expandafter \@secondoftwo
 \fi
}%
\providecommand \natexlab [1]{#1}%
\providecommand \enquote  [1]{``#1''}%
\providecommand \bibnamefont  [1]{#1}%
\providecommand \bibfnamefont [1]{#1}%
\providecommand \citenamefont [1]{#1}%
\providecommand \href@noop [0]{\@secondoftwo}%
\providecommand \href [0]{\begingroup \@sanitize@url \@href}%
\providecommand \@href[1]{\@@startlink{#1}\@@href}%
\providecommand \@@href[1]{\endgroup#1\@@endlink}%
\providecommand \@sanitize@url [0]{\catcode `\\12\catcode `\$12\catcode
  `\&12\catcode `\#12\catcode `\^12\catcode `\_12\catcode `\%12\relax}%
\providecommand \@@startlink[1]{}%
\providecommand \@@endlink[0]{}%
\providecommand \url  [0]{\begingroup\@sanitize@url \@url }%
\providecommand \@url [1]{\endgroup\@href {#1}{\urlprefix }}%
\providecommand \urlprefix  [0]{URL }%
\providecommand \Eprint [0]{\href }%
\providecommand \doibase [0]{https://doi.org/}%
\providecommand \selectlanguage [0]{\@gobble}%
\providecommand \bibinfo  [0]{\@secondoftwo}%
\providecommand \bibfield  [0]{\@secondoftwo}%
\providecommand \translation [1]{[#1]}%
\providecommand \BibitemOpen [0]{}%
\providecommand \bibitemStop [0]{}%
\providecommand \bibitemNoStop [0]{.\EOS\space}%
\providecommand \EOS [0]{\spacefactor3000\relax}%
\providecommand \BibitemShut  [1]{\csname bibitem#1\endcsname}%
\let\auto@bib@innerbib\@empty
\bibitem [{\citenamefont {\text{Letter} of Heisenberg~to
  Peierls~(1930)}(1985)}]{H1}%
  \BibitemOpen
  \bibfield  {author} {\bibinfo {author} {\bibnamefont {\text{Letter} of
  Heisenberg~to Peierls~(1930)}},\ }\href@noop {} {\emph {\bibinfo {title}
  {Wolfgang Pauli, Scientific Correspondence}}},\ Vol.~\bibinfo {volume} {2}\
  (\bibinfo  {publisher} {Springer Berlin, Heidelberg},\ \bibinfo {year}
  {1985})\BibitemShut {NoStop}%
\bibitem [{\citenamefont {Peierls}(1933)}]{peierls33}%
  \BibitemOpen
  \bibfield  {author} {\bibinfo {author} {\bibfnamefont {R.}~\bibnamefont
  {Peierls}},\ }\href {https://doi.org/https://doi.org/10.1007/BF01342591}
  {\bibfield  {journal} {\bibinfo  {journal} {Z. Rev.}\ }\textbf {\bibinfo
  {volume} {80}},\ \bibinfo {pages} {763–791} (\bibinfo {year}
  {1933})}\BibitemShut {NoStop}%
\bibitem [{\citenamefont {Snyder}(1947)}]{Snyder:1946qz}%
  \BibitemOpen
  \bibfield  {author} {\bibinfo {author} {\bibfnamefont {H.~S.}\ \bibnamefont
  {Snyder}},\ }\href {https://doi.org/10.1103/PhysRev.71.38} {\bibfield
  {journal} {\bibinfo  {journal} {Phys. Rev.}\ }\textbf {\bibinfo {volume}
  {71}},\ \bibinfo {pages} {38} (\bibinfo {year} {1947})}\BibitemShut {NoStop}%
\bibitem [{\citenamefont {Yang}(1947)}]{Yang47}%
  \BibitemOpen
  \bibfield  {author} {\bibinfo {author} {\bibfnamefont {C.~N.}\ \bibnamefont
  {Yang}},\ }\href {https://doi.org/10.1103/PhysRev.72.874} {\bibfield
  {journal} {\bibinfo  {journal} {Phys. Rev.}\ }\textbf {\bibinfo {volume}
  {72}},\ \bibinfo {pages} {874} (\bibinfo {year} {1947})}\BibitemShut
  {NoStop}%
\bibitem [{\citenamefont {Connes}\ \emph {et~al.}(1998)\citenamefont {Connes},
  \citenamefont {Douglas},\ and\ \citenamefont {Schwarz}}]{Connes:1997cr}%
  \BibitemOpen
  \bibfield  {author} {\bibinfo {author} {\bibfnamefont {A.}~\bibnamefont
  {Connes}}, \bibinfo {author} {\bibfnamefont {M.~R.}\ \bibnamefont
  {Douglas}},\ and\ \bibinfo {author} {\bibfnamefont {A.~S.}\ \bibnamefont
  {Schwarz}},\ }\href {https://doi.org/10.1088/1126-6708/1998/02/003}
  {\bibfield  {journal} {\bibinfo  {journal} {JHEP}\ }\textbf {\bibinfo
  {volume} {02}},\ \bibinfo {pages} {003 (1998)}}\BibitemShut {NoStop}%
\bibitem [{\citenamefont {Banks}\ \emph {et~al.}(1997)\citenamefont {Banks},
  \citenamefont {Fischler}, \citenamefont {Shenker},\ and\ \citenamefont
  {Susskind}}]{Banks:1996vh}%
  \BibitemOpen
  \bibfield  {author} {\bibinfo {author} {\bibfnamefont {T.}~\bibnamefont
  {Banks}}, \bibinfo {author} {\bibfnamefont {W.}~\bibnamefont {Fischler}},
  \bibinfo {author} {\bibfnamefont {S.~H.}\ \bibnamefont {Shenker}},\ and\
  \bibinfo {author} {\bibfnamefont {L.}~\bibnamefont {Susskind}},\ }\href
  {https://doi.org/10.1103/PhysRevD.55.5112} {\bibfield  {journal} {\bibinfo
  {journal} {Phys. Rev. D}\ }\textbf {\bibinfo {volume} {55}},\ \bibinfo
  {pages} {5112} (\bibinfo {year} {1997})}\BibitemShut {NoStop}%
\bibitem [{\citenamefont {Ishibashi}\ \emph {et~al.}(1997)\citenamefont
  {Ishibashi}, \citenamefont {Kawai}, \citenamefont {Kitazawa},\ and\
  \citenamefont {Tsuchiya}}]{Ishibashi:1996xs}%
  \BibitemOpen
  \bibfield  {author} {\bibinfo {author} {\bibfnamefont {N.}~\bibnamefont
  {Ishibashi}}, \bibinfo {author} {\bibfnamefont {H.}~\bibnamefont {Kawai}},
  \bibinfo {author} {\bibfnamefont {Y.}~\bibnamefont {Kitazawa}},\ and\
  \bibinfo {author} {\bibfnamefont {A.}~\bibnamefont {Tsuchiya}},\ }\href
  {https://doi.org/10.1016/S0550-3213(97)00290-3} {\bibfield  {journal}
  {\bibinfo  {journal} {Nucl. Phys. B}\ }\textbf {\bibinfo {volume} {498}},\
  \bibinfo {pages} {467} (\bibinfo {year} {1997})}\BibitemShut {NoStop}%
\bibitem [{\citenamefont {Seiberg}\ and\ \citenamefont
  {Witten}(1999)}]{Seiberg:1999vs}%
  \BibitemOpen
  \bibfield  {author} {\bibinfo {author} {\bibfnamefont {N.}~\bibnamefont
  {Seiberg}}\ and\ \bibinfo {author} {\bibfnamefont {E.}~\bibnamefont
  {Witten}},\ }\href {https://doi.org/10.1088/1126-6708/1999/09/032} {\bibfield
   {journal} {\bibinfo  {journal} {JHEP}\ }\textbf {\bibinfo {volume} {09}},\
  \bibinfo {pages} {032 (1999)}}\BibitemShut {NoStop}%
\bibitem [{\citenamefont {Balachandran}\ \emph {et~al.}(2006)\citenamefont
  {Balachandran}, \citenamefont {Mangano}, \citenamefont {Pinzul},\ and\
  \citenamefont {Vaidya}}]{moyalplaneBalachandran:2005eb}%
  \BibitemOpen
  \bibfield  {author} {\bibinfo {author} {\bibfnamefont {A.~P.}\ \bibnamefont
  {Balachandran}}, \bibinfo {author} {\bibfnamefont {G.}~\bibnamefont
  {Mangano}}, \bibinfo {author} {\bibfnamefont {A.}~\bibnamefont {Pinzul}},\
  and\ \bibinfo {author} {\bibfnamefont {S.}~\bibnamefont {Vaidya}},\ }\href
  {https://doi.org/10.1142/S0217751X06031764} {\bibfield  {journal} {\bibinfo
  {journal} {Int. J. Mod. Phys. A}\ }\textbf {\bibinfo {volume} {21}},\
  \bibinfo {pages} {3111} (\bibinfo {year} {2006})}\BibitemShut {NoStop}%
\bibitem [{\citenamefont {Akofor}\ \emph {et~al.}(2008)\citenamefont {Akofor},
  \citenamefont {Balachandran},\ and\ \citenamefont
  {Joseph}}]{moyalplane3Akofor:2008ae}%
  \BibitemOpen
  \bibfield  {author} {\bibinfo {author} {\bibfnamefont {E.}~\bibnamefont
  {Akofor}}, \bibinfo {author} {\bibfnamefont {A.~P.}\ \bibnamefont
  {Balachandran}},\ and\ \bibinfo {author} {\bibfnamefont {A.}~\bibnamefont
  {Joseph}},\ }\href {https://doi.org/10.1142/S0217751X08040317} {\bibfield
  {journal} {\bibinfo  {journal} {Int. J. Mod. Phys. A}\ }\textbf {\bibinfo
  {volume} {23}},\ \bibinfo {pages} {1637} (\bibinfo {year}
  {2008})}\BibitemShut {NoStop}%
\bibitem [{\citenamefont {Drinfeld}(1986)}]{Drinfeld:1986in}%
  \BibitemOpen
  \bibfield  {author} {\bibinfo {author} {\bibfnamefont {V.~G.}\ \bibnamefont
  {Drinfeld}},\ }\href {https://doi.org/10.1007/BF01247086} {\bibfield
  {journal} {\bibinfo  {journal} {Zap. Nauchn. Semin.}\ }\textbf {\bibinfo
  {volume} {155}},\ \bibinfo {pages} {18} (\bibinfo {year} {1986})}\BibitemShut
  {NoStop}%
\bibitem [{\citenamefont {Chaichian}\ \emph {et~al.}(2004)\citenamefont
  {Chaichian}, \citenamefont {Kulish}, \citenamefont {Nishijima},\ and\
  \citenamefont {Tureanu}}]{theta1Chaichian:2004za}%
  \BibitemOpen
  \bibfield  {author} {\bibinfo {author} {\bibfnamefont {M.}~\bibnamefont
  {Chaichian}}, \bibinfo {author} {\bibfnamefont {P.~P.}\ \bibnamefont
  {Kulish}}, \bibinfo {author} {\bibfnamefont {K.}~\bibnamefont {Nishijima}},\
  and\ \bibinfo {author} {\bibfnamefont {A.}~\bibnamefont {Tureanu}},\ }\href
  {https://doi.org/10.1016/j.physletb.2004.10.045} {\bibfield  {journal}
  {\bibinfo  {journal} {Phys. Lett. B}\ }\textbf {\bibinfo {volume} {604}},\
  \bibinfo {pages} {98} (\bibinfo {year} {2004})}\BibitemShut {NoStop}%
\bibitem [{\citenamefont {Addazi}\ and\ \citenamefont
  {Marcian\`o}(2020)}]{theta2Addazi:2018jmt}%
  \BibitemOpen
  \bibfield  {author} {\bibinfo {author} {\bibfnamefont {A.}~\bibnamefont
  {Addazi}}\ and\ \bibinfo {author} {\bibfnamefont {A.}~\bibnamefont
  {Marcian\`o}},\ }\href {https://doi.org/10.1142/S0217751X20420038} {\bibfield
   {journal} {\bibinfo  {journal} {Int. J. Mod. Phys. A}\ }\textbf {\bibinfo
  {volume} {35}},\ \bibinfo {pages} {2042003} (\bibinfo {year}
  {2020})}\BibitemShut {NoStop}%
\bibitem [{\citenamefont {Piscicchia}\ \emph {et~al.}(2022)\citenamefont
  {Piscicchia} \emph {et~al.}}]{strongestpepv}%
  \BibitemOpen
  \bibfield  {author} {\bibinfo {author} {\bibfnamefont {K.}~\bibnamefont
  {Piscicchia}} \emph {et~al.},\ }\href
  {https://doi.org/10.1103/PhysRevLett.129.131301} {\bibfield  {journal}
  {\bibinfo  {journal} {Phys. Rev. Lett.}\ }\textbf {\bibinfo {volume} {129}},\
  \bibinfo {pages} {131301} (\bibinfo {year} {2022})}\BibitemShut {NoStop}%
\bibitem [{\citenamefont {Minwalla}\ \emph {et~al.}(2000)\citenamefont
  {Minwalla}, \citenamefont {Van~Raamsdonk},\ and\ \citenamefont
  {Seiberg}}]{Minwalla_2000}%
  \BibitemOpen
  \bibfield  {author} {\bibinfo {author} {\bibfnamefont {S.}~\bibnamefont
  {Minwalla}}, \bibinfo {author} {\bibfnamefont {M.}~\bibnamefont
  {Van~Raamsdonk}},\ and\ \bibinfo {author} {\bibfnamefont {N.}~\bibnamefont
  {Seiberg}},\ }\href {https://doi.org/10.1088/1126-6708/2000/02/020}
  {\bibfield  {journal} {\bibinfo  {journal} {JHEP}\ }\textbf {\bibinfo
  {volume} {02}},\ \bibinfo {pages} {020 (2000)}}\BibitemShut {NoStop}%
\bibitem [{\citenamefont {Craig}\ and\ \citenamefont
  {Koren}(2020)}]{Craig_2020}%
  \BibitemOpen
  \bibfield  {author} {\bibinfo {author} {\bibfnamefont {N.}~\bibnamefont
  {Craig}}\ and\ \bibinfo {author} {\bibfnamefont {S.}~\bibnamefont {Koren}},\
  }\href {https://doi.org/10.1007/JHEP03(2020)037} {\bibfield  {journal}
  {\bibinfo  {journal} {JHEP}\ }\textbf {\bibinfo {volume} {03}},\ \bibinfo
  {pages} {037 (2020)}}\BibitemShut {NoStop}%
\bibitem [{\citenamefont {Matusis}\ \emph {et~al.}(2000)\citenamefont
  {Matusis}, \citenamefont {Susskind},\ and\ \citenamefont {Toumbas}}]{UVIR3}%
  \BibitemOpen
  \bibfield  {author} {\bibinfo {author} {\bibfnamefont {A.}~\bibnamefont
  {Matusis}}, \bibinfo {author} {\bibfnamefont {L.}~\bibnamefont {Susskind}},\
  and\ \bibinfo {author} {\bibfnamefont {N.}~\bibnamefont {Toumbas}},\ }\href
  {https://doi.org/10.1088/1126-6708/2000/12/002} {\bibfield  {journal}
  {\bibinfo  {journal} {JHEP}\ }\textbf {\bibinfo {volume} {12}},\ \bibinfo
  {pages} {002 (2000)}}\BibitemShut {NoStop}%
\bibitem [{\citenamefont {Balachandran}\ \emph {et~al.}(2010)\citenamefont
  {Balachandran}, \citenamefont {Joseph},\ and\ \citenamefont
  {Padmanabhan}}]{pepv2Balachandran:2010xk}%
  \BibitemOpen
  \bibfield  {author} {\bibinfo {author} {\bibfnamefont {A.~P.}\ \bibnamefont
  {Balachandran}}, \bibinfo {author} {\bibfnamefont {A.}~\bibnamefont
  {Joseph}},\ and\ \bibinfo {author} {\bibfnamefont {P.}~\bibnamefont
  {Padmanabhan}},\ }\href {https://doi.org/10.1103/PhysRevLett.105.051601}
  {\bibfield  {journal} {\bibinfo  {journal} {Phys. Rev. Lett.}\ }\textbf
  {\bibinfo {volume} {105}},\ \bibinfo {pages} {051601} (\bibinfo {year}
  {2010})}\BibitemShut {NoStop}%
\bibitem [{\citenamefont {Balachandran}\ and\ \citenamefont
  {Padmanabhan}(2010)}]{paperBalachandran:2010wq}%
  \BibitemOpen
  \bibfield  {author} {\bibinfo {author} {\bibfnamefont {A.~P.}\ \bibnamefont
  {Balachandran}}\ and\ \bibinfo {author} {\bibfnamefont {P.}~\bibnamefont
  {Padmanabhan}},\ }\href {https://doi.org/10.1007/JHEP12(2010)001} {\bibfield
  {journal} {\bibinfo  {journal} {JHEP}\ }\textbf {\bibinfo {volume} {12}},\
  \bibinfo {pages} {001 (2010)}}\BibitemShut {NoStop}%
\bibitem [{\citenamefont {Messiah}\ and\ \citenamefont
  {Greenberg}(1964)}]{PEPVst2}%
  \BibitemOpen
  \bibfield  {author} {\bibinfo {author} {\bibfnamefont {A.~M.~L.}\
  \bibnamefont {Messiah}}\ and\ \bibinfo {author} {\bibfnamefont {O.~W.}\
  \bibnamefont {Greenberg}},\ }\href {https://doi.org/10.1103/PhysRev.136.B248}
  {\bibfield  {journal} {\bibinfo  {journal} {Phys. Rev.}\ }\textbf {\bibinfo
  {volume} {136}},\ \bibinfo {pages} {B248} (\bibinfo {year}
  {1964})}\BibitemShut {NoStop}%
\bibitem [{\citenamefont {Gavrin}\ \emph {et~al.}(1988)\citenamefont {Gavrin},
  \citenamefont {Ignatiev},\ and\ \citenamefont {Kuzmin}}]{PEPVst3}%
  \BibitemOpen
  \bibfield  {author} {\bibinfo {author} {\bibfnamefont {V.~N.}\ \bibnamefont
  {Gavrin}}, \bibinfo {author} {\bibfnamefont {A.~Y.}\ \bibnamefont
  {Ignatiev}},\ and\ \bibinfo {author} {\bibfnamefont {V.}~\bibnamefont
  {Kuzmin}},\ }\href@noop {} {\bibfield  {journal} {\bibinfo  {journal} {{Phy.
  Lett. B}}\ }\textbf {\bibinfo {volume} {206}},\ \bibinfo {pages} {343}
  (\bibinfo {year} {1988})}\BibitemShut {NoStop}%
\bibitem [{\citenamefont {Mohapatra}(1990)}]{PEPVst4}%
  \BibitemOpen
  \bibfield  {author} {\bibinfo {author} {\bibfnamefont {R.~N.}\ \bibnamefont
  {Mohapatra}},\ }\href@noop {} {\bibfield  {journal} {\bibinfo  {journal}
  {{Phy. Lett. B}}\ }\textbf {\bibinfo {volume} {242}},\ \bibinfo {pages} {407}
  (\bibinfo {year} {1990})}\BibitemShut {NoStop}%
\bibitem [{\citenamefont {Greenberg}\ and\ \citenamefont
  {Mohapatra}(1987)}]{pepv1Greenberg:1987ih}%
  \BibitemOpen
  \bibfield  {author} {\bibinfo {author} {\bibfnamefont {O.~W.}\ \bibnamefont
  {Greenberg}}\ and\ \bibinfo {author} {\bibfnamefont {R.~N.}\ \bibnamefont
  {Mohapatra}},\ }\href {https://doi.org/10.1103/PhysRevLett.59.2507}
  {\bibfield  {journal} {\bibinfo  {journal} {Phys. Rev. Lett.}\ }\textbf
  {\bibinfo {volume} {59}},\ \bibinfo {pages} {2507} (\bibinfo {year}
  {1987})},\ \bibinfo {note} {[Erratum: Phys.Rev.Lett. 61, 1432
  (1988)]}\BibitemShut {NoStop}%
\bibitem [{\citenamefont {Ramberg}\ and\ \citenamefont
  {Snow}(1990)}]{pepv5Ramberg:1988iu}%
  \BibitemOpen
  \bibfield  {author} {\bibinfo {author} {\bibfnamefont {E.}~\bibnamefont
  {Ramberg}}\ and\ \bibinfo {author} {\bibfnamefont {G.~A.}\ \bibnamefont
  {Snow}},\ }\href {https://doi.org/10.1016/0370-2693(90)91762-Z} {\bibfield
  {journal} {\bibinfo  {journal} {Phys. Lett. B}\ }\textbf {\bibinfo {volume}
  {238}},\ \bibinfo {pages} {438} (\bibinfo {year} {1990})}\BibitemShut
  {NoStop}%
\bibitem [{\citenamefont {Bartalucci}\ \emph {et~al.}(2006)\citenamefont
  {Bartalucci} \emph {et~al.}}]{pepv6VIP:2006gvw}%
  \BibitemOpen
  \bibfield  {author} {\bibinfo {author} {\bibfnamefont {S.}~\bibnamefont
  {Bartalucci}} \emph {et~al.} (\bibinfo {collaboration} {VIP Collaboration}),\
  }\href {https://doi.org/10.1016/j.physletb.2006.07.054} {\bibfield  {journal}
  {\bibinfo  {journal} {Phys. Lett. B}\ }\textbf {\bibinfo {volume} {641}},\
  \bibinfo {pages} {18} (\bibinfo {year} {2006})}\BibitemShut {NoStop}%
\bibitem [{\citenamefont {Back}\ \emph {et~al.}(2004)\citenamefont {Back} \emph
  {et~al.}}]{pepv7Borexino:2004hfc}%
  \BibitemOpen
  \bibfield  {author} {\bibinfo {author} {\bibfnamefont {H.~O.}\ \bibnamefont
  {Back}} \emph {et~al.} (\bibinfo {collaboration} {Borexino Collaboration}),\
  }\href {https://doi.org/10.1140/epjc/s2004-01991-1} {\bibfield  {journal}
  {\bibinfo  {journal} {Eur. Phys. J. C}\ }\textbf {\bibinfo {volume} {37}},\
  \bibinfo {pages} {421} (\bibinfo {year} {2004})}\BibitemShut {NoStop}%
\bibitem [{\citenamefont {Hazewinkel}\ \emph {et~al.}(2010)\citenamefont
  {Hazewinkel} \emph {et~al.}}]{Hazewinkel10}%
  \BibitemOpen
  \bibfield  {author} {\bibinfo {author} {\bibfnamefont {M.}~\bibnamefont
  {Hazewinkel}} \emph {et~al.},\ }\href@noop {} {\emph {\bibinfo {title}
  {Algebras, rings and modules: Lie algebras and Hopf algebras}}},\
  Vol.~\bibinfo {volume} {3}\ (\bibinfo  {publisher} {American Mathematical
  Soc. RI},\ \bibinfo {year} {2010})\BibitemShut {NoStop}%
\bibitem [{\citenamefont {Majid}(2011)}]{Majid1995}%
  \BibitemOpen
  \bibfield  {author} {\bibinfo {author} {\bibfnamefont {S.}~\bibnamefont
  {Majid}},\ }\href@noop {} {\emph {\bibinfo {title} {{Foundations of quantum
  group theory}}}}\ (\bibinfo  {publisher} {Cambridge University Press,
  Cambridge},\ \bibinfo {year} {2011})\BibitemShut {NoStop}%
\bibitem [{\citenamefont {Weinzierl}(2004)}]{Weinzierl03}%
  \BibitemOpen
  \bibfield  {author} {\bibinfo {author} {\bibfnamefont {S.}~\bibnamefont
  {Weinzierl}},\ }\href {https://doi.org/10.1140/epjcd/s2003-03-1001-y}
  {\bibfield  {journal} {\bibinfo  {journal} {Eur. Phys. J. C}\ }\textbf
  {\bibinfo {volume} {33}},\ \bibinfo {pages} {S871} (\bibinfo {year}
  {2004})}\BibitemShut {NoStop}%
\bibitem [{\citenamefont {Thompson}\ \emph {et~al.}(2009)\citenamefont
  {Thompson}, \citenamefont {Vaughan} \emph {et~al.}}]{t2}%
  \BibitemOpen
  \bibfield  {author} {\bibinfo {author} {\bibfnamefont {A.~C.}\ \bibnamefont
  {Thompson}}, \bibinfo {author} {\bibfnamefont {D.}~\bibnamefont {Vaughan}},
  \emph {et~al.},\ }\href@noop {} {\emph {\bibinfo {title} {X-ray data
  booklet}}}\ (\bibinfo  {publisher} {Lawrence Berkeley National Laboratory,
  University of California Berkeley, CA},\ \bibinfo {year} {2009})\BibitemShut
  {NoStop}%
\bibitem [{\citenamefont {Arnaud}\ \emph {et~al.}(2018)\citenamefont {Arnaud}
  \emph {et~al.}}]{ng2}%
  \BibitemOpen
  \bibfield  {author} {\bibinfo {author} {\bibfnamefont {Q.}~\bibnamefont
  {Arnaud}} \emph {et~al.} (\bibinfo {collaboration} {NEWS-G Collaboration}),\
  }\href {https://doi.org/10.1016/j.astropartphys.2017.10.009} {\bibfield
  {journal} {\bibinfo  {journal} {Astropart. Phys.}\ }\textbf {\bibinfo
  {volume} {97}},\ \bibinfo {pages} {54} (\bibinfo {year} {2018})}\BibitemShut
  {NoStop}%
\bibitem [{\citenamefont {Aguilar-Arevalo}\ \emph {et~al.}(2022)\citenamefont
  {Aguilar-Arevalo} \emph {et~al.}}]{dami2}%
  \BibitemOpen
  \bibfield  {author} {\bibinfo {author} {\bibfnamefont {A.}~\bibnamefont
  {Aguilar-Arevalo}} \emph {et~al.} (\bibinfo {collaboration} {DAMIC
  Collaboration}),\ }\href {https://doi.org/10.1103/PhysRevD.105.062003}
  {\bibfield  {journal} {\bibinfo  {journal} {Phys. Rev. D}\ }\textbf {\bibinfo
  {volume} {105}},\ \bibinfo {pages} {062003} (\bibinfo {year}
  {2022})}\BibitemShut {NoStop}%
\bibitem [{\citenamefont {Agnes}\ \emph {et~al.}()\citenamefont {Agnes} \emph
  {et~al.}}]{dk2}%
  \BibitemOpen
  \bibfield  {author} {\bibinfo {author} {\bibfnamefont {P.}~\bibnamefont
  {Agnes}} \emph {et~al.} (\bibinfo {collaboration} {DarkSide-50
  Collaboration}),\ }\href@noop {} {\bibinfo  {journal} {arXiv:2207.11966,
  (2022)}\ }\BibitemShut {NoStop}%
\bibitem [{\citenamefont {Angloher}\ \emph {et~al.}()\citenamefont {Angloher}
  \emph {et~al.}}]{cr4}%
  \BibitemOpen
\bibfield  {journal} {  }\bibfield  {author} {\bibinfo {author} {\bibfnamefont
  {G.}~\bibnamefont {Angloher}} \emph {et~al.} (\bibinfo {collaboration}
  {CRESST-II Collaboration}),\ }\href@noop {} {\bibinfo  {journal}
  {arXiv:1701.08157, (2017)}\ }\BibitemShut {NoStop}%
\bibitem [{\citenamefont {Arnquist}\ \emph {et~al.}()\citenamefont {Arnquist}
  \emph {et~al.}}]{mjd2}%
  \BibitemOpen
\bibfield  {journal} {  }\bibfield  {author} {\bibinfo {author} {\bibfnamefont
  {I.~J.}\ \bibnamefont {Arnquist}} \emph {et~al.} (\bibinfo {collaboration}
  {Majorana Collaboration}),\ }\href@noop {} {\bibinfo  {journal}
  {arXiv:2206.10638, (2022)}\ }\BibitemShut {NoStop}%
\bibitem [{\citenamefont {Bernabei}\ \emph {et~al.}(2008)\citenamefont
  {Bernabei} \emph {et~al.}}]{dama2}%
  \BibitemOpen
\bibfield  {journal} {  }\bibfield  {author} {\bibinfo {author} {\bibfnamefont
  {R.}~\bibnamefont {Bernabei}} \emph {et~al.} (\bibinfo {collaboration} {DAMA
  Collaboration}),\ }\href {https://doi.org/10.1016/j.nima.2008.04.082}
  {\bibfield  {journal} {\bibinfo  {journal} {Nucl. Instrum. Meth. A}\ }\textbf
  {\bibinfo {volume} {592}},\ \bibinfo {pages} {297} (\bibinfo {year}
  {2008})}\BibitemShut {NoStop}%
\bibitem [{\citenamefont {Aprile}\ \emph {et~al.}(2022)\citenamefont {Aprile}
  \emph {et~al.}}]{xe4}%
  \BibitemOpen
  \bibfield  {author} {\bibinfo {author} {\bibfnamefont {E.}~\bibnamefont
  {Aprile}} \emph {et~al.} (\bibinfo {collaboration} {XENON Collaboration}),\
  }\href {https://doi.org/10.1103/PhysRevLett.129.161805} {\bibfield  {journal}
  {\bibinfo  {journal} {Phys. Rev. Lett.}\ }\textbf {\bibinfo {volume} {129}},\
  \bibinfo {pages} {161805} (\bibinfo {year} {2022})}\BibitemShut {NoStop}%
\bibitem [{\citenamefont {Bernabei}\ \emph {et~al.}(2020)\citenamefont
  {Bernabei} \emph {et~al.}}]{dama1}%
  \BibitemOpen
  \bibfield  {author} {\bibinfo {author} {\bibfnamefont {R.}~\bibnamefont
  {Bernabei}} \emph {et~al.},\ }\href@noop {} {\bibfield  {journal} {\bibinfo
  {journal} {Progress in Particle and Nuclear Physics}\ }\textbf {\bibinfo
  {volume} {114}},\ \bibinfo {pages} {103810} (\bibinfo {year}
  {2020})}\BibitemShut {NoStop}%
\bibitem [{\citenamefont {Angloher}\ \emph {et~al.}(2014)\citenamefont
  {Angloher} \emph {et~al.}}]{cr3}%
  \BibitemOpen
  \bibfield  {author} {\bibinfo {author} {\bibfnamefont {G.}~\bibnamefont
  {Angloher}} \emph {et~al.} (\bibinfo {collaboration} {CRESST-II
  Collaboration}),\ }\href {https://doi.org/10.1140/epjc/s10052-014-3184-9}
  {\bibfield  {journal} {\bibinfo  {journal} {Eur. Phys. J. C}\ }\textbf
  {\bibinfo {volume} {74}},\ \bibinfo {pages} {3184} (\bibinfo {year}
  {2014})}\BibitemShut {NoStop}%
\bibitem [{\citenamefont {Szydagis}\ \emph {et~al.}(2021)\citenamefont
  {Szydagis}, \citenamefont {Levy}, \citenamefont {Blockinger}, \citenamefont
  {Kamaha}, \citenamefont {Parveen},\ and\ \citenamefont {Rischbieter}}]{xe5}%
  \BibitemOpen
  \bibfield  {author} {\bibinfo {author} {\bibfnamefont {M.}~\bibnamefont
  {Szydagis}}, \bibinfo {author} {\bibfnamefont {C.}~\bibnamefont {Levy}},
  \bibinfo {author} {\bibfnamefont {G.~M.}\ \bibnamefont {Blockinger}},
  \bibinfo {author} {\bibfnamefont {A.}~\bibnamefont {Kamaha}}, \bibinfo
  {author} {\bibfnamefont {N.}~\bibnamefont {Parveen}},\ and\ \bibinfo {author}
  {\bibfnamefont {G.~R.~C.}\ \bibnamefont {Rischbieter}},\ }\href
  {https://doi.org/10.1103/PhysRevD.103.012002} {\bibfield  {journal} {\bibinfo
   {journal} {Phys. Rev. D}\ }\textbf {\bibinfo {volume} {103}},\ \bibinfo
  {pages} {012002} (\bibinfo {year} {2021})}\BibitemShut {NoStop}%
\bibitem [{\citenamefont {Agnes}\ \emph {et~al.}(2021)\citenamefont {Agnes}
  \emph {et~al.}}]{dk4}%
  \BibitemOpen
  \bibfield  {author} {\bibinfo {author} {\bibfnamefont {P.}~\bibnamefont
  {Agnes}} \emph {et~al.} (\bibinfo {collaboration} {DarkSide Collaboration}),\
  }\href {https://doi.org/10.1103/PhysRevD.104.082005} {\bibfield  {journal}
  {\bibinfo  {journal} {Phys. Rev. D}\ }\textbf {\bibinfo {volume} {104}},\
  \bibinfo {pages} {082005} (\bibinfo {year} {2021})}\BibitemShut {NoStop}%
\bibitem [{\citenamefont {Fano}(1947)}]{t4}%
  \BibitemOpen
  \bibfield  {author} {\bibinfo {author} {\bibfnamefont {U.}~\bibnamefont
  {Fano}},\ }\href {https://doi.org/10.1103/PhysRev.72.26} {\bibfield
  {journal} {\bibinfo  {journal} {Phys. Rev.}\ }\textbf {\bibinfo {volume}
  {72}},\ \bibinfo {pages} {26} (\bibinfo {year} {1947})}\BibitemShut {NoStop}%
\bibitem [{\citenamefont {Feldman}\ and\ \citenamefont {Cousins}(1998)}]{t3}%
  \BibitemOpen
  \bibfield  {author} {\bibinfo {author} {\bibfnamefont {G.~J.}\ \bibnamefont
  {Feldman}}\ and\ \bibinfo {author} {\bibfnamefont {R.~D.}\ \bibnamefont
  {Cousins}},\ }\href {https://doi.org/10.1103/PhysRevD.57.3873} {\bibfield
  {journal} {\bibinfo  {journal} {Phys. Rev. D}\ }\textbf {\bibinfo {volume}
  {57}},\ \bibinfo {pages} {3873} (\bibinfo {year} {1998})}\BibitemShut
  {NoStop}%
\bibitem [{\citenamefont {Scofield}(1974)}]{t5}%
  \BibitemOpen
  \bibfield  {author} {\bibinfo {author} {\bibfnamefont {J.~H.}\ \bibnamefont
  {Scofield}},\ }\href {https://doi.org/10.1016/S0092-640X(74)80019-7}
  {\bibfield  {journal} {\bibinfo  {journal} {Atom. Data Nucl. Data Tabl.}\
  }\textbf {\bibinfo {volume} {14}},\ \bibinfo {pages} {121} (\bibinfo {year}
  {1974})}\BibitemShut {NoStop}%
\bibitem [{\citenamefont {Bingham}\ \emph {et~al.}(2006)\citenamefont
  {Bingham}, \citenamefont {Mendon{\c c}a},\ and\ \citenamefont
  {Wang}}]{report}%
  \BibitemOpen
  \bibfield  {author} {\bibinfo {author} {\bibfnamefont {R.}~\bibnamefont
  {Bingham}}, \bibinfo {author} {\bibfnamefont {T.}~\bibnamefont {Mendon{\c
  c}a}},\ and\ \bibinfo {author} {\bibfnamefont {C.}~\bibnamefont {Wang}},\
  }\href@noop {} {\bibfield  {journal} {\bibinfo  {journal} {CERN Courier}\ ,\
  \bibinfo {pages} {25}} (\bibinfo {year} {2006})}\BibitemShut {NoStop}%
\bibitem [{\citenamefont {Ellis}\ \emph {et~al.}(1984)\citenamefont {Ellis},
  \citenamefont {Hagelin}, \citenamefont {Nanopoulos},\ and\ \citenamefont
  {Srednicki}}]{timeq1}%
  \BibitemOpen
  \bibfield  {author} {\bibinfo {author} {\bibfnamefont {J.~R.}\ \bibnamefont
  {Ellis}}, \bibinfo {author} {\bibfnamefont {J.~S.}\ \bibnamefont {Hagelin}},
  \bibinfo {author} {\bibfnamefont {D.~V.}\ \bibnamefont {Nanopoulos}},\ and\
  \bibinfo {author} {\bibfnamefont {M.}~\bibnamefont {Srednicki}},\ }\href
  {https://doi.org/10.1016/0550-3213(84)90053-1} {\bibfield  {journal}
  {\bibinfo  {journal} {Nucl. Phys. B}\ }\textbf {\bibinfo {volume} {241}},\
  \bibinfo {pages} {381} (\bibinfo {year} {1984})}\BibitemShut {NoStop}%
\end{thebibliography}%

\end{document}